\documentclass[sigconf]{acmart}

\usepackage[ruled, linesnumbered]{algorithm2e}
\usepackage{enumitem}
\usepackage{multirow}
\usepackage{subfig}

\newcommand{\bs}[1]{\mathbf{#1}}
\newcommand{\model}[1]{\textbf{#1}}
\newcommand\ie{\textit{i.e.}}
\newcommand\eg{\textit{e.g.}}

\DeclareMathOperator*{\argmin}{arg\,min}

\AtBeginDocument{%
  }

\copyrightyear{2026}
\acmYear{2026}
\setcopyright{cc}
\setcctype{by}
\acmConference[CIKM '26]{Proceedings of the 35th ACM International Conference on Information and Knowledge Management}{November 07--11, 2026}{Rome, Italy}
\acmBooktitle{Proceedings of the 35th ACM International Conference on Information and Knowledge Management (CIKM '26), November 07--11, 2026, Rome, Italy}
\acmDOI{10.1145/3799682.3840900}
\acmISBN{979-8-4007-2539-5/2026/11}

\begin{document}
\title{RQ-Reg: A Residual-Quantization-Based Framework for Continuous Value Prediction in Recommender Systems}

\author{Runpeng Cui}
\authornote{Corresponding author: Runpeng Cui <cuirunpeng@gmail.com>.}
\affiliation{
  \institution{Kuaishou Technology}
  \city{Beijing}
  \country{China}
}
\email{cuirunpeng@kuaishou.com}

\author{Zhipeng Sun}
\affiliation{
  \institution{Kuaishou Technology}
  \city{Beijing}
  \country{China}
}
\email{sunzhipeng03@kuaishou.com}

\author{Chi Lu}
\affiliation{
  \institution{Kuaishou Technology}
  \city{Beijing}
  \country{China}
}
\email{luchi@kuaishou.com}

\author{Peng Jiang}
\affiliation{
 \institution{Kuaishou Technology}
 \city{Beijing}
 \country{China}
}
\email{jiangpeng@kuaishou.com}

\renewcommand{\shortauthors}{Runpeng Cui, Zhipeng Sun, Chi Lu, and Peng Jiang}

\begin{abstract}
Predicting continuous values such as watch-time and gross merchandise value (GMV) is a core problem in industrial recommendation systems.
Its inherent difficulty stems from the highly complex and long-tailed distributions of the target signals, which are hard to model accurately.
Existing regression methods typically rely on fixed parametric assumptions on the target distribution: overly simple assumptions underfit real-world data, whereas more intricate ones tend to sacrifice scalability and generalization.

To address these limitations, we propose a sequence modeling framework based on residual quantization (RQ),
in which the target continuous value is decomposed into a sequence of quantization codes that represent progressively finer approximations.
The model autoregressively predicts these codes from coarse to fine granularity, with each step refining the residual error left by the previous one.
To further improve the quality of the learned representations, we introduce an ordinal-aware representation learning objective that aligns the RQ code embedding space with the ordinal structure of target values,
thereby yielding continuous representations of quantization codes and more accurate predictions.
We conduct comprehensive experiments on public benchmarks for watch-time and lifetime value (LTV) prediction,
together with a large-scale online A/B test for GMV prediction on an industrial short-video recommendation platform.
Across all settings, the proposed method shows competitive performance among existing state-of-the-art approaches and generalizes well across diverse continuous value prediction scenarios.
\end{abstract}



\maketitle

\section{Introduction}
Recommendation systems have made substantial progress in recent years in matching users with items of interest.
With the rapid growth of online platforms ranging from short-video applications to large-scale e-commerce services,
predicting continuous-valued signals has become increasingly important.
Key engagement and business metrics, such as watch-time~\cite{lin2023tree,sun2024cread,yang2025swat}, lifetime value (LTV)~\cite{weng2024optdist},
and gross merchandise value (GMV)~\cite{xin2019multi},
are crucial continuous signals that capture a user's interests, engagement and commercial potential.
For example, on short-video platforms such as TikTok and YouTube Shorts,
watch-time serves as a key behavioral signal reflecting user interest and engagement.
Sustained viewing not only indicates user satisfaction but also drives
substantial business value through mechanisms such as targeted advertising and in-app e-commerce.
Accurately predicting a user's watch-time on candidate videos has therefore
become a critical task for improving content relevance, boosting user engagement, and maximizing platform effectiveness.

Continuous value prediction in recommendation systems is naturally formulated as a regression task~\cite{covington2016deep,zhan2022deconfounding},
in which the objective is typically to minimize the deviation between predicted and ground-truth values.
However, since the target distribution is wide-ranging, complex, and often heavily skewed (see Fig.~\ref{fig:gmv}),
standard regression losses such as $\ell_1$ or mean squared error (MSE) tend to yield suboptimal results on long-tailed instances.
To mitigate this issue, classification-based approaches~\cite{lin2023tree,sun2024cread}
discretize the target distribution into predefined intervals and then train classifiers over the resulting buckets.
The central difficulty here lies in balancing classification accuracy against the reconstruction error introduced by discretization~\cite{sun2024cread}:
wide bucket intervals simplify classification but incur larger quantization errors,
whereas narrow intervals suffer from data sparsity and class imbalance that degrade classifier effectiveness.

In contrast to classification, generative modeling approaches~\cite{wang2019deep,zhang2023out,weng2024optdist,yang2025swat}
assume a specific parametric form for the target distribution and fit its parameters via maximum likelihood estimation (MLE).
For such methods, prediction performance is fundamentally constrained by the appropriateness of the distributional assumption:
overly simplified forms cannot adequately model real-world complexity and lead to inferior predictions,
while more intricate assumptions often suffer from poor scalability and generalization.

In light of these limitations, we propose \textbf{RQ-Reg}, a novel residual quantization (RQ)-based sequence learning framework for continuous value prediction.
Unlike prior classification methods that rely on a fixed, global discretization scheme,
our method learns a coarse-to-fine decomposition of the target value via RQ.
Specifically, the target value is represented as a sum of ordered quantization codes,
allowing the model to first capture the coarse magnitude and then progressively refine the prediction through recursive residual estimation.
This design retains the training stability of coarse discretization while substantially reducing the quantization error,
effectively resolving the core trade-off in prior approaches.
Furthermore, we introduce a representation learning objective that explicitly aligns the embedding structure with the ranking of target values,
enabling the model to capture continuous relationships in the embedding space and further improve prediction accuracy.

To the best of our knowledge, this is the first work to integrate RQ with sequence learning for continuous value prediction in recommender systems,
offering both performance gains and methodological generality.
Our main contributions are as follows:
\begin{itemize}[leftmargin=*]
  \item We propose a novel RQ-based decomposition for continuous value prediction,
  which represents the target value as a sum of ordered quantization codes
  and predicts them recursively under a sequence learning architecture.
  \item We introduce a representation learning objective that aligns the embedding space
  with the ranking in the target space, enabling the model to capture continuous representations
  of quantization codes and further enhancing prediction performance.
  \item Extensive experiments on public benchmarks and a large-scale online short-video platform
  show that our framework compares favorably with state-of-the-art (SOTA) continuous value prediction methods
  and generalizes to a broad range of continuous prediction challenges.
  We further conduct detailed ablation studies to quantify the contribution of each key component.
\end{itemize}

\section{Related Work}
\subsection{Regression by Classification}
Classification-based approaches have been successfully applied to regression tasks
such as watch-time~\cite{lin2023tree,sun2024cread,chen2025personalized} and LTV prediction~\cite{chamberlain2017customer,li2022billion} in recommendation scenarios.
These methods typically discretize the continuous target into predefined groups and train classifiers over the resulting buckets.
Among them, ordinal regression~\cite{gutierrez2015ordinal,li2022billion,sun2024cread} is one of the most widely adopted formulations,
as it explicitly leverages the inherent ordering of the categories within the classification framework.
However, most classification-based methods rely on a fixed, predefined discretization scheme,
which forces a trade-off between classification accuracy and discretization error~\cite{sun2024cread}
and often fails to capture the complexity of real-world target distributions.

\begin{figure}
  \centering
  \includegraphics[width=8.4cm]{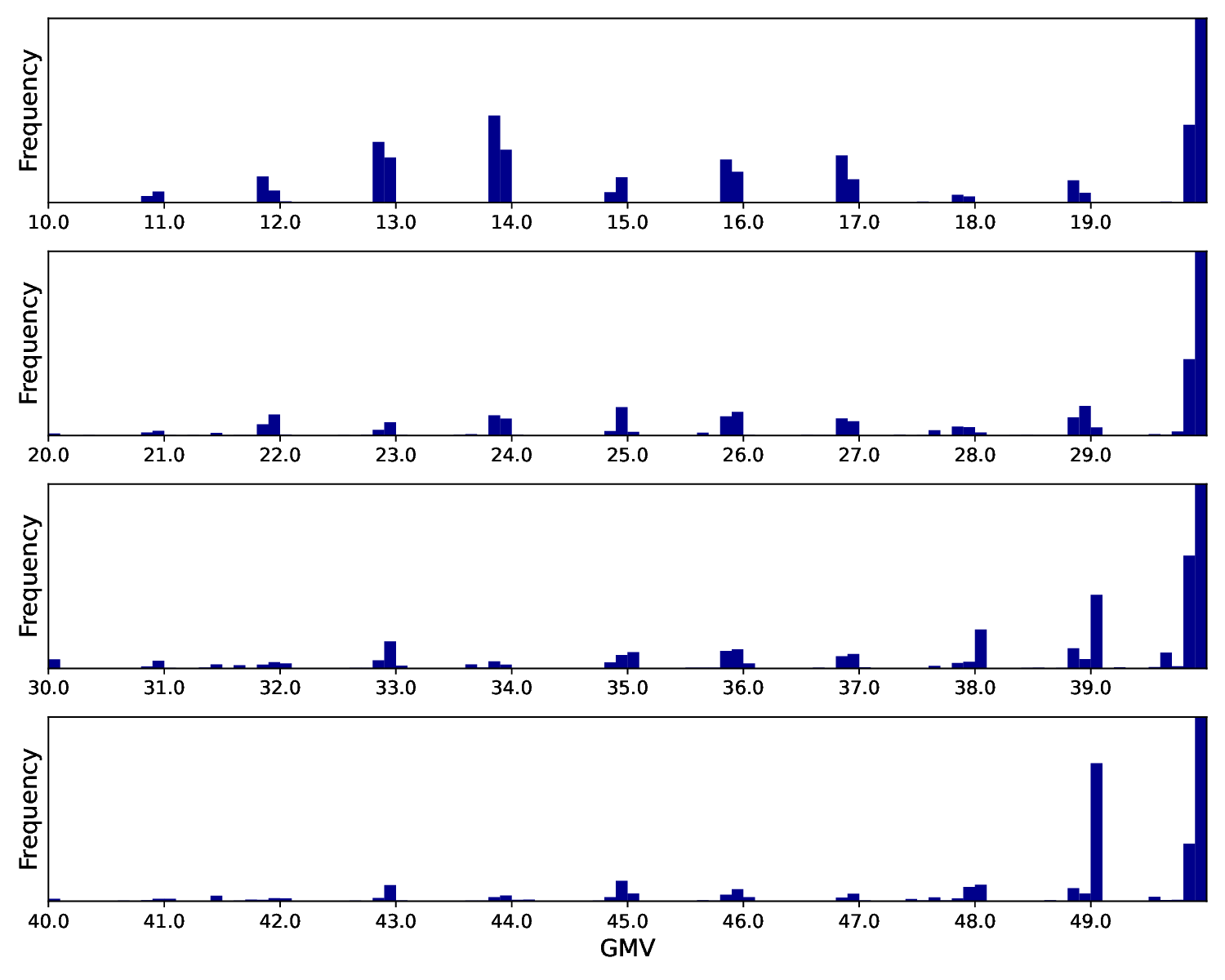}
  \caption{GMV distribution in a large-scale online short-video commercial scenario.
  From top to bottom, each subfigure shows the GMV frequency distribution over the value ranges 10--20, 20--30, 30--40, and 40--50.}
  \label{fig:gmv}
\end{figure}

\subsection{Regression by Generative Approaches}
Generative methods~\cite{covington2016deep,wang2019deep,zhang2023out,weng2024optdist,yang2025swat}
typically assume a specific parametric form for the target distribution and fit its parameters via MLE.
Some early works~\cite{covington2016deep,zhan2022deconfounding,zheng2022dvr}
make simple distributional assumptions and adopt regression losses such as the $\ell_1$ loss or MSE to train the model.
However, these simple regression formulations tend to be suboptimal on long-tailed instances
due to the complexity and skewness of the underlying target distribution.

More recently, generative approaches have achieved notable success
on regression tasks in recommendation systems~\cite{wang2019deep,zhao2023uncovering,zhang2023out,zhao2024counteracting,weng2024optdist,zhao2025multi}.
For instance, Weng et al.~\cite{weng2024optdist} model the customer LTV distribution as
a mixture of zero-inflated lognormal distributions~\cite{wang2019deep},
and use distribution learning and selection modules to address the learning problem.
For generative methods, prediction performance is fundamentally constrained by the appropriateness of the distributional assumption.
Most existing approaches rely on simplifying assumptions that limit their ability to capture the inherent complexity of the target distribution,
whereas overly complicated assumptions on distributions can lead to poor scalability and limited generalization of the method.

In contrast, our RQ-Reg method decomposes the continuous target via RQ encoding
and reformulates prediction as a coarse-to-fine sequence learning problem.
Experiments show that RQ-Reg can be readily applied to a variety of continuous value prediction tasks in recommendation scenarios,
demonstrating its generality across a broad range of challenges.

We also note that Ma et al.~\cite{ma2024generative} recently propose a generative regression (GR) method
that formulates watch-time prediction as a sequence generation task.
Our method differs from GR in several important aspects.
First, GR uses a heuristic algorithm with additional monotonicity constraints to decompose the target into variable-length codes,
which complicates training.
In contrast, we are the first to introduce RQ to continuous value prediction, achieving lower reconstruction error and more accurate prediction.
Second, GR uses a Transformer-based model whose complexity scales quadratically with the code length,
whereas RQ-Reg attains linear complexity in the code length and thus incurs substantially lower computational overhead.
Moreover, we introduce a representation learning objective~\cite{zha2023rank} that explicitly aligns the embedding structure with the ordinal nature of target values,
allowing the model to capture intrinsic sample relationships and further improve predictive accuracy.

\section{Method}
\subsection{Notations}
Let $\mathcal{X}=\left\{\left(\bs{x}^i, y^i\right)\right\}_{i=1}^N$ denote the training set,
where $\bs{x}^i\in\mathbb{R}^d$ is the feature vector of the $i$-th sample and $y^i\in\mathcal{Y}\subset\mathbb{R}_+$ is the corresponding label.
The goal of the prediction model is to produce an accurate estimate $\hat{y}^i$ of the ground-truth continuous value $y^i$.
Let $\mathcal{C}=(\mathcal{C}_1,\cdots,\mathcal{C}_L)$ denote a hierarchical codebook consisting of $L$ levels,
where each level $\mathcal{C}_l\subset\mathbb{R}$ contains $K$ scalar codes.
For brevity, we omit the sample superscript $i$ whenever it is unambiguous.

\subsection{Residual Quantization}
Inspired by residual quantization (RQ)~\cite{lee2022autoregressive},
which transforms feature vectors into discrete codes,
we propose to decompose the continuous target value $y$ and discretize it into coarse-to-fine estimates via RQ.

We develop a hierarchical K-means clustering algorithm for RQ (RQ K-means)~\cite{luo2025qarm}
to construct an RQ codebook $\mathcal{C}$ with $L$ levels.
Let $\mathcal{R}_0=\mathcal{Y}$ denote the set of $0$-th residuals of the target values in the training set.
For $l=1,\cdots, L$, the $l$-th level codebook is obtained as
\begin{equation}
\mathcal{C}_l = \text{K-means}(\mathcal{R}_{l-1}, K),
\end{equation}
where $\mathcal{C}_l=\{c_{l,k}\}_{k=1}^K$ is the set of K-means cluster centroids.
Given a residual $r_{l-1}\in\mathcal{R}_{l-1}$, we denote by $Q(r_{l-1};\mathcal{C}_l)$ the quantization code at the $l$-th level,
which is the centroid closest to $r_{l-1}$:
\begin{equation}
  Q(r_{l-1};\mathcal{C}_l)=c_{l,\kappa},
\end{equation}
where
\begin{equation}
  \kappa = \argmin_k\ \left(r_{l-1} - c_{l,k}\right)^2.
\end{equation}
The next-level residual $r_l$ is then computed as
\begin{equation}
r_l=r_{l-1} - Q(r_{l-1};\mathcal{C}_l).
\end{equation}
By recursively applying the above procedure for $l=1,\cdots,L$, we obtain the full codebook $\mathcal{C}$.
The complete RQ K-means procedure is summarized in Algorithm~\ref{al:rq_kmeans}.

\begin{algorithm}
\caption{RQ K-means for continuous value discretization}\label{al:rq_kmeans}
\KwIn{Continuous value set $\mathcal{Y}$}
\KwOut{Codebook $\mathcal{C}$}
Initialize the $0$-th residual set $\mathcal{R}_0=\mathcal{Y}$\;
\For{$l=1$ \KwTo $L$} {
  $\mathcal{C}_l=\text{K-means}(\mathcal{R}_{l-1},K)$\;
  $\mathcal{R}_{l}=\left\{r:r = u - Q(u;\mathcal{C}_l),\ u\in\mathcal{R}_{l-1}\right\}$\;
}
Return $\mathcal{C}=\left(\mathcal{C}_1,\cdots,\mathcal{C}_L\right)$\;
\end{algorithm}

We let $q_l=Q(r_{l-1};\mathcal{C}_l)$ denote the $l$-th quantization code,
and define $y_{(l)}=\sum_{i=1}^l q_i$ as the partial sum of the first $l$ quantization codes.
Starting from the $0$-th residual $r_0=y$, the continuous target value $y\in\mathcal{Y}$ can be recursively quantized as
\begin{equation}
\label{eq:quant}
y=\left(\sum_{l=1}^L q_l\right)+r_L = y_{(L)}+r_L,
\end{equation}
where $r_L$ is the quantization error at the $L$-th level.
At each level, the nearest centroid $q_l$ is selected to approximate the current residual, thereby minimizing the per-level quantization error.
As $l$ increases from $1$ to $L$, the partial sum $y_{(l)}$ yields increasingly fine-grained approximations of $y$.

Note that K-means clustering minimizes the within-cluster variance of samples,
\ie, the sum of squared quantization errors.
By applying K-means recursively to the residuals at each level,
our RQ K-means procedure substantially reduces the sum of quantization errors
over all samples in the training set.
Following~\cite{lee2022autoregressive},
in practice we use a single shared codebook $\mathcal{C}=\bigcup_{l=1}^L\mathcal{C}_l$ across all quantization levels,
and we denote the codebook size as $|\mathcal{C}|=L\times K$.

\subsection{RQ Modeling for Continuous Value Prediction}
With the help of the quantization procedure of RQ K-means,
the target value $y$ can be decomposed into a sequence of $L$ quantization codes $\{q_l\}_{l=1}^L$.
Since the quantization errors are minimized by RQ K-means,
we have $y=\left(\sum_{l=1}^L q_l\right) + r_L$ with $r_L$ being a small residual.
The continuous value prediction task can therefore be reformulated as a sequence learning problem
where the model produces sequential predictions for the quantization code at each step.

\subsubsection{RQ codes modeling.}
We decompose the target value $y$ into the RQ code sequence $\{q_l\}_{l=1}^L$ as described above.
As the codes are generated autoregressively, the joint probability $p\left(q_1,\cdots,q_L|\bs{x}\right)$ is factorized as
\begin{equation}
  \label{eq:prod}
  p\left(q_1,\cdots,q_L|\bs{x}\right)=\prod_{l=1}^L p\left(q_l|\bs{x},q_{<l}\right).
\end{equation}

\begin{figure}
  \centering
  \includegraphics[width=8.3cm]{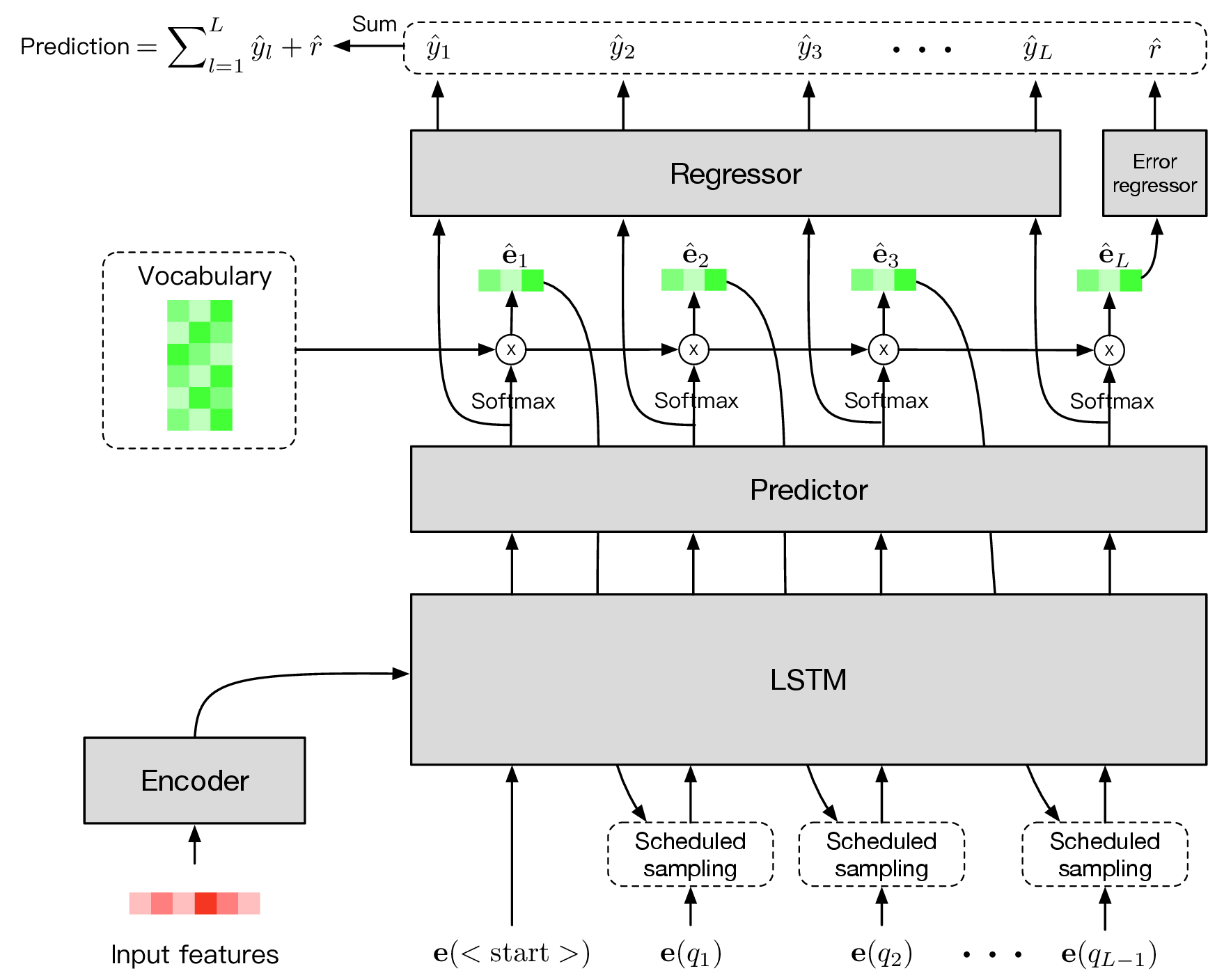}
  \caption{Architecture of the proposed RQ-Reg method.}
  \label{fig:diagram}
\end{figure}

\subsubsection{Architecture.}
We design an RNN-based sequence learning architecture to model the quantization code sequence,
and adopt a Long Short-Term Memory (LSTM)~\cite{hochreiter1997long} network as the backbone.
Following common practice in natural language processing, we set the initial code $q_0$ to a special token $\text{<start>}$.
We build an embedding table $\bs{E}\in\mathbb{R}^{|\mathcal{V}|\times D}$ to map every code
in the vocabulary $\mathcal{V}=\mathcal{C}\cup\{\text{<start>}\}$ to a representation in the embedding space.
To be specific, each code $q$ is represented as a $D$-dimensional embedding $\bs{e}(q)\in\mathbb{R}^D$,
which is the corresponding row of $\bs{E}$.

Given the concatenated input feature vector $\bs{x}$,
we use a multi-layer perceptron (MLP) $f_{\text{enc}}$ to produce the initial hidden state and memory cell of the LSTM:
\begin{equation}
  \left[\bs{h}_0;\bs{c}_0\right]=f_{\text{enc}}(\bs{x}),
\end{equation}
where $\left[\cdot;\cdot\right]$ denotes concatenation.

At step $l$ ($1\le l\le L$), we first obtain the representation $\bs{e}\left(q_{l-1}\right)$ of code $q_{l-1}$ from $\bs{E}$,
and feed it into the LSTM:
\begin{equation}
  \bs{h}_l,\ \bs{c}_l = \text{LSTM}\left(\bs{e}(q_{l-1}),\bs{h}_{l-1},\bs{c}_{l-1}\right).
\end{equation}
During training, instead of always using the ground-truth code embedding $\bs{e}(q_{l-1})$ as input,
we adopt a scheduled sampling scheme~\cite{bengio2015scheduled} that stochastically
replaces it with the model's own prediction $\hat{\bs{e}}_{l-1}$ (defined in Eq.~(\ref{eq:mix_emb})) from the previous step.
This mitigates exposure bias by progressively bridging the gap between training and inference,
as elaborated in Section~\ref{sssec:training}.

The model predicts the conditional distribution $p\left(q|\bs{x},q_{<l}\right)$ of the next code as
\begin{equation}
  p\left(q|\bs{x},q_{<l}\right)=\text{softmax}\left(\bs{z}_l\right),
\end{equation}
where $\bs{z}_l=f_\text{pred}(\bs{h}_l)$ and $f_\text{pred}$ is a predictor that transforms the LSTM output into
the logits $\bs{z}_l\in\mathbb{R}^{|\mathcal{V}|}$ over the RQ code vocabulary.
To produce an accurate prediction of the RQ-code value at step $l$,
we further introduce an MLP regressor $f_\text{reg}:\mathbb{R}^{|\mathcal{V}|}\to \mathbb{R}$,
which learns a mapping from the logits $\bs{z}_l$ to the quantization code value:
\begin{equation}
  \hat{y}_l=f_\text{reg}\left(\bs{z}_l\right),
\end{equation}
where $\hat{y}_l$ is the prediction of the $l$-th quantization code value.

After all the predictions $\hat{y}_1,\cdots,\hat{y}_L$ of RQ codes have been generated,
we use an additional MLP $f_\text{err}:\mathbb{R}^D\to \mathbb{R}$ to predict the residual quantization error: $\hat{r}=f_\text{err}\left(\hat{\bs{e}}_L\right)$.

Given input features $\bs{x}$ and the initial code $q_0$,
our architecture runs recursively to predict the RQ-code values $\hat{y}_1,\cdots,\hat{y}_L$ and the error estimate $\hat{r}$,
and therefore provides
\begin{equation}
  \hat{y}=\hat{y}_{(L)} + \hat{r}
\end{equation}
as the final estimate of the target value $y$,
where $\hat{y}_{(l)}=\sum_{i=1}^l \hat{y}_i$ denotes the partial sum of the RQ-code predictions up to step $l$.
The overall architecture of RQ-Reg is illustrated in Fig.~\ref{fig:diagram}.

\subsection{Objective Function}
\subsubsection{RQ code generation.}
As described in Eq.~(\ref{eq:prod}),
the sequence learning model recursively outputs the probability $p\left(q=q_l|\bs{x},q_{<l}\right)$ at each step $l$.
The architecture is trained to minimize the negative log-likelihood (NLL) loss:
\begin{equation}
  \mathcal{L}_\text{gen}=-\frac{1}{N}\sum_{(\bs{x},y)\in\mathcal{X}}\sum_{l=1}^L\log p\left(q=q_l|\bs{x},q_{<l}\right).
\end{equation}

\subsubsection{Regression.}
We use the deviations between predictions and ground-truth values as the regression objective.
Following prior work~\cite{sun2024cread}, we employ the Huber loss~\cite{huber1992robust}, defined as
\begin{equation}
  \ell_\delta(u,v)=
  \begin{cases}
    \frac{1}{2}(u - v)^2, & \text{if } |u - v| \leq \delta, \\
    \delta \cdot|u - v| - \frac{1}{2}\delta^2, & \text{otherwise},
    \end{cases}
\end{equation}
with a threshold $\delta$, to formulate the regression objective as
\begin{equation}
  \mathcal{L}_\text{reg}=\frac{1}{N}\sum_{(\bs{x},y)\in\mathcal{X}}\left[\ell_\delta(y, \hat{y}) + \sum_{l=1}^L\ell_\delta\left(y_{(l)}, \hat{y}_{(l)}\right)\right].
\end{equation}
The objective $\mathcal{L}_\text{reg}$ encourages the model to produce accurate predictions
both for the final target value $y$ and for the coarse-to-fine quantized values $y_{(l)}$ at each level.

\subsubsection{Continuous representation learning.}
Given the embedding table $\bs{E}$ and the distribution $p\left(q|\bs{x},q_{<l}\right)$ predicted at step $l$,
we represent the quantization code at the $l$-th level in the embedding space as
\begin{equation}
  \hat{\bs{e}}_l = \sum_{c\in\mathcal{V}}\left[p\left(q=c|\bs{x},q_{<l}\right)\cdot \bs{e}(c)\right],
  \label{eq:mix_emb}
\end{equation}
which can be interpreted as the expectation of the $l$-th-level code representations with respect to the conditional distribution.
We then define $\hat{\bs{e}}=\left[\hat{\bs{e}}_1;\cdots;\hat{\bs{e}}_L\right]$ as the representation of the entire quantization code sequence.
To shape this representation space so that it reflects the ordinal structure of the target values,
we introduce the Rank-N-Contrast (RnC) loss~\cite{zha2023rank}.
Let $\text{sim}(\cdot,\cdot)$ denote a similarity measure between two vectors, and $\tau$ denote the temperature parameter.
The RnC loss for sample $i$ over all the other samples in a batch $\mathcal{B}$ is defined as
\begin{equation}
  \ell_\text{rnc}^{(i)}=-\frac{1}{|\mathcal{B}|-1}\sum_{j\in\mathcal{B},\ j\neq i}\log\frac{\exp\left(\text{sim}(\hat{\bs{e}}^i, \hat{\bs{e}}^j) / \tau\right)}{\sum_{k\in\mathcal{S}^{i,j}}\exp\left(\text{sim}(\hat{\bs{e}}^i,\hat{\bs{e}}^k) / \tau\right)},
\end{equation}
where $\mathcal{S}^{i,j}=\left\{k:k\neq i,|y^i-y^k|\ge|y^i-y^j|\right\}$ is the set of samples
of higher ranks than sample $j$ in terms of label distance with respect to sample $i$.
By minimizing $\ell_\text{rnc}^{(i)}$, the model is encouraged to
align the order of the code-sequence representations with their corresponding order in the label space with respect to anchor sample $i$.
We then have
$\mathcal{L}_\text{rnc}=\frac{1}{N}\sum_{\left(\bs{x}^i,y^i\right)\in\mathcal{X}}\ell_\text{rnc}^{(i)}$,
which calculates the contrastive term over all training samples,
enforcing the entire embedding space to be ordered according to the order in the label space.

In summary, the overall objective function for RQ-Reg is
\begin{equation}
  \mathcal{L}=\mathcal{L}_\text{gen} + \alpha\mathcal{L}_\text{reg} + \beta\mathcal{L}_\text{rnc},
\end{equation}
where $\alpha$ and $\beta$ are hyperparameters that balance the components of the overall objective.

\subsection{Optimization and Inference}

\subsubsection{Training.}
\label{sssec:training}
At the training stage, we adopt a scheduled sampling mechanism~\cite{bengio2015scheduled}, which is a curriculum learning strategy,
to improve the training stability and effectiveness of sequential RQ-code prediction.
At the early stages of training, sampling from the model outputs would yield random tokens since the model is not yet well trained,
which can lead to slow convergence, so selecting the true previous token more often should help.
Towards the end of training, by contrast, we sample from the model more often,
as this corresponds to the true inference situation,
and a well-trained model is expected to sample reasonable tokens autonomously.

Specifically, instead of always feeding the ground-truth RQ-code embedding
$\bs{e}(q_{l})$ into the LSTM at decoding step $l+1$,
we adopt a stochastic scheduled sampling scheme.
With probability $1-p$, the ground-truth embedding $\bs{e}(q_{l})$
is replaced by the predicted embedding $\hat{\bs{e}}_l$
derived from the model's own output distribution via Eq.~(\ref{eq:mix_emb}).
As $p$ decreases over the course of training,
this interpolation between teacher forcing and model self-feeding
facilitates a smooth transition from fully supervised guidance to autonomous sequence generation.
Let $t$ denote the training epoch.
We adopt the inverse sigmoid decay approach~\cite{bengio2015scheduled} as:
\begin{equation}
  p(t)=\text{sigmoid}\left(-k\left(t-t_0\right)\right),
  \label{eq:inv_sigmoid}
\end{equation}
where $k>0$ and $t_0$ are hyperparameters that control the shape of the decay curve.

\subsubsection{Inference.}
During inference, the RQ-Reg model runs in a fully autoregressive manner.
At each step $l$, the model consumes the predicted embedding $\hat{\bs{e}}_{l-1}$
from the previous step to generate the conditional distribution
$p\left(q|\bs{x},q_{<l}\right)$ and the corresponding predicted value $\hat{y}_l$.
In addition, we obtain the predicted RQ error $\hat{r}$ at step $L$.
When all $L$ predictions for codes and the error estimate $\hat{r}$ have been generated,
the final prediction is computed as $\hat{y}=\left(\sum_{l=1}^L \hat{y}_l\right) + \hat{r}$.
This fully autoregressive inference procedure ensures consistency
between the predicted RQ code sequence and the resulting continuous label.

\section{Experiments}
In this section, we evaluate the proposed RQ-Reg on continuous value prediction tasks.
We describe the offline experiment setup on public benchmarks in Section~\ref{sub:setup},
and present and analyze the corresponding results in Section~\ref{sub:exp}.
To further demonstrate effectiveness in real-world scenarios,
Section~\ref{sub:online} reports an online experiment on a large-scale short-video recommendation platform.

\subsection{Experiment Setup}
\label{sub:setup}

\subsubsection{Datasets.}
We conduct offline experiments on two continuous value prediction tasks in recommendation scenarios, \ie, watch-time and LTV prediction,
each evaluated on two public datasets:
KuaiRec~\cite{gao2022kuairec} and CIKM16\footnote{\url{https://competitions.codalab.org/competitions/11161}} for watch-time prediction,
and Criteo-SSC~\cite{tallis2018reacting} and Kaggle\footnote{\url{https://www.kaggle.com/c/acquire-valued-shoppers-challenge}} for LTV prediction.
\begin{itemize}[leftmargin=*]
  \item \textbf{KuaiRec} is a watch-time prediction dataset collected from the Kuaishou app, a short-video platform,
  containing 7{,}176 users, 10{,}728 items, and 12{,}530{,}806 impressions.
  \item \textbf{CIKM16} is a watch-time prediction dataset collected from logs of an e-commerce search engine,
  containing 122{,}991 items and 310{,}302 sessions.
  \item \textbf{Criteo-SSC} is an LTV prediction dataset collected from Criteo Predictive Search (CPS), a search marketing platform,
  containing 15{,}995{,}633 samples with a positive sample ratio of 7.20\%.
  \item \textbf{Kaggle} is an LTV prediction dataset built from the shopping histories of a large set of shoppers,
  containing over 300{,}000 shoppers and 805{,}753 samples with a positive sample ratio of 90.12\%.
\end{itemize}

\begin{table}
  \caption{Performance comparison on watch-time prediction datasets}
  \begin{center}
  \begin{tabular}{lcccc}
  \toprule
  \multirow{2}{*}{Method} & \multicolumn{2}{c}{KuaiRec} & \multicolumn{2}{c}{CIKM16} \\
  \cmidrule{2-5}
    & MAE$^\downarrow$ & XAUC$^\uparrow$ & MAE$^\downarrow$ & XAUC$^\uparrow$ \\\midrule
  WLR~\cite{covington2016deep} & 6.047 & 0.525 & 0.998 & 0.672 \\
  D2Q~\cite{zhan2022deconfounding} & 5.426 & 0.565 & 0.899 & 0.661 \\
  TPM~\cite{lin2023tree} & 4.741 & 0.599 & 0.884 & 0.676 \\
  CREAD~\cite{sun2024cread} & 3.215 & 0.601 & 0.865 & 0.678 \\
  SWaT~\cite{yang2025swat} & 3.363 & 0.609 & 0.847 & 0.683 \\
  GR~\cite{ma2024generative} & 3.196 & 0.614 & 0.815 & 0.691 \\
  \midrule
  RQ-Reg & \textbf{3.189} & \textbf{0.616} & \textbf{0.811} & \textbf{0.695} \\
  \bottomrule
  \end{tabular}
  \end{center}
  \label{tab:wtp}
\end{table}

\subsubsection{Implementation details.}
For all datasets, we use a single-layer LSTM backbone with hidden size $512$ for sequence learning.
The encoder $f_\text{enc}$, predictor $f_\text{pred}$, and regressors $f_\text{reg}$ and $f_\text{err}$ are all implemented as two-layer MLPs with ReLU activations.
For the residual quantization configuration, we set the number of RQ levels to $L=3$ and the number of codes per level to $K=48$.
The parameters of RQ-Reg are optimized with the AdamW~\cite{loshchilov2017decoupled} optimizer for a maximum of 50 epochs,
using a learning rate of $5.0\times 10^{-4}$.
All RQ-Reg results reported in this paper are averaged over three independent runs,
and baseline results are quoted from the respective publications~\cite{ma2024generative,weng2024optdist,yang2025swat}.
Additional implementation details are provided in Appendix~\ref{appen:detail}.
Our source code for RQ-Reg is available at \url{https://github.com/rpcui/RQ-Reg}.

To ensure a fair comparison, for the watch-time task we follow the setup of~\cite{lin2023tree,ma2024generative} and partition CIKM16 into training and test sets with an 8:2 split,
and adopt the same feature processing pipeline as~\cite{lin2023tree}.
For the LTV prediction task, the Criteo-SSC and Kaggle datasets are split into training, validation, and test sets in a 7:1:2 ratio,
following the setup of~\cite{weng2024optdist}.

\begin{table*}
  \setlength{\tabcolsep}{9pt}
  \caption{Performance comparison on LTV prediction datasets}
  \begin{center}
  \begin{tabular}{clccccc}
  \toprule
  Dataset & Method & MAE$^\downarrow$ & Gini~(+)$^\uparrow$ & SRCC~(+)$^\uparrow$ & Gini$^\uparrow$ & SRCC$^\uparrow$ \\
  \midrule
  \multirow{7}{*}{Criteo-SSC} & Two-stage~\cite{drachen2018or} & 21.719 & 0.2204 & 0.2565 & 0.5278 & 0.2386 \\
  & MTL-MSE~\cite{ma2018entire} & 21.190 & 0.4340 & 0.3663 & 0.6330 & 0.2478 \\
  & ZILN~\cite{wang2019deep} & 20.880 & 0.4426 & 0.3874 & 0.6338 & 0.2434 \\
  & MDME~\cite{li2022billion} & 16.598 & 0.2297 & 0.2952 & 0.4383 & 0.2269 \\
  & MDAN~\cite{liu2024mdan} & 20.030 & 0.4128 & 0.3521 & 0.6209 & 0.2470 \\
  & OptDist~\cite{weng2024optdist} & \textbf{15.784} & 0.4428 & 0.3903 & \textbf{0.6437} & 0.2505 \\
  \cmidrule{2-7}
  & RQ-Reg & 16.329 & \textbf{0.4468} & \textbf{0.4166} & 0.6141 & \textbf{0.2523} \\
  \midrule
  \multirow{7}{*}{Kaggle} & Two-stage~\cite{drachen2018or} & 74.782 & 0.5505 & 0.4596 & 0.5498 & 0.4313 \\
  & MTL-MSE~\cite{ma2018entire} & 74.065 & 0.5349 & 0.4328 & 0.5503 & 0.4329 \\
  & ZILN~\cite{wang2019deep} & 72.528 & 0.6627 & 0.5303 & 0.6693 & 0.5239 \\
  & MDME~\cite{li2022billion} & 72.900 & 0.6213 & 0.5289 & 0.6305 & 0.5163 \\
  & MDAN~\cite{liu2024mdan} & 73.940 & 0.6680 & 0.4567 & 0.6648 & 0.4367 \\
  & OptDist~\cite{weng2024optdist} & 70.929 & 0.6715 & 0.5346 & 0.6814 & 0.5249 \\
  \cmidrule{2-7}
  & RQ-Reg & \textbf{59.340} & \textbf{0.7201} & \textbf{0.5358} & \textbf{0.7235} & \textbf{0.5318} \\
  \bottomrule
  \end{tabular}
  \end{center}
  \label{tab:ltv}
\end{table*}

\subsubsection{Metrics.}
For the watch-time prediction task, following~\cite{lin2023tree,ma2024generative},
we report the mean absolute error (MAE) and XAUC~\cite{zhan2022deconfounding} as performance metrics:
MAE measures the absolute deviation between predicted and ground-truth values,
and XAUC measures the consistency between the ranking order of the predicted values and that of the ground-truth values.
Specifically, XAUC is defined as
\begin{equation}
  \text{XAUC} = \frac{1}{|\mathcal{P}|}\sum_{(i,j)\in\mathcal{P}}\bs{1}\left[\left(y^i - y^j\right)\left(\hat{y}^i - \hat{y}^j\right)>0\right],
\end{equation}
where $\bs{1}[\cdot]$ is the indicator function, $\hat{y}^i$ is the predicted value for sample $i$, and $\mathcal{P}$ is the set of sample pairs in the test set.

For the LTV task, following prior work~\cite{wang2019deep,yang2023feature,weng2024optdist},
we adopt the normalized Gini coefficient (Gini)
and the Spearman rank correlation coefficient (SRCC) to evaluate the overall ranking performance over all samples.
Following~\cite{weng2024optdist}, we additionally evaluate the models on positive samples
separately using the same two metrics, denoted Gini~(+) and SRCC~(+) respectively,
to assess the ability of each method to discriminate among positive LTV samples.

\begin{table}
  \caption{Ablation study on watch-time prediction datasets}
  \begin{center}
  \begin{tabular}{lcccc}
  \toprule
  \multirow{2}{*}{Method} & \multicolumn{2}{c}{KuaiRec} & \multicolumn{2}{c}{CIKM16} \\
  \cmidrule{2-5}
   & MAE$^\downarrow$ & XAUC$^\uparrow$ & MAE$^\downarrow$ & XAUC$^\uparrow$ \\\midrule
   RQ-Reg & \textbf{3.189} & \textbf{0.616} & \textbf{0.811} & \textbf{0.695} \\
   \midrule
  w/o $\mathcal{L}_\text{gen}$ & 3.199 & 0.613 & 0.821 & 0.691 \\
  w/o $\mathcal{L}_\text{rnc}$ & 3.195 & 0.614 & 0.819 & 0.692 \\\midrule
  w/o $f_\text{err}$ & 3.191 & 0.614 & 0.813 & 0.694 \\
  w/o $f_\text{reg}$ + $f_\text{err}$ & 3.200 & 0.613 & 0.815 & 0.694 \\\midrule
  w/o SS~($p=1$) & 3.240 & 0.604 & 0.880 & 0.688 \\
  w/o SS~($p=0$) & 3.197 & 0.613 & 0.823 & 0.691 \\
  \bottomrule
  \end{tabular}
  \end{center}
  \label{tab:ab_main}
\end{table}

\subsection{Results on Public Datasets}
\label{sub:exp}
\subsubsection{Main results.}
Table~\ref{tab:wtp} reports the results on the watch-time prediction benchmarks in comparison with SOTA approaches,
including WLR~\cite{covington2016deep}, D2Q~\cite{zhan2022deconfounding}, TPM~\cite{lin2023tree},
CREAD~\cite{sun2024cread}, SWaT~\cite{yang2025swat}, and GR~\cite{ma2024generative},
covering both regression accuracy and ranking consistency.
RQ-Reg consistently achieves the best performance across all metrics and datasets.
The improvements over TPM~\cite{lin2023tree} and CREAD~\cite{sun2024cread}
demonstrate the advantages of our approach over conventional classification-based regression methods.
RQ-Reg also yields clear gains over regression-based baselines~\cite{covington2016deep,zhan2022deconfounding,yang2025swat},
confirming the benefit of our sequential prediction framework.
Furthermore, RQ-Reg consistently outperforms GR~\cite{ma2024generative},
achieving relative improvements of $0.33\%$ and $0.58\%$ in XAUC on KuaiRec and CIKM16 respectively,
together with relative reductions of $0.22\%$ and $0.49\%$ in MAE on the two datasets.
Specifically, on KuaiRec, RQ-Reg attains MAE $=3.189\pm 0.001$ and XAUC $=0.6155\pm 0.0002$ (vs.~$3.196$ and $0.614$ for GR);
on CIKM16, MAE $=0.811\pm 0.001$ and XAUC $=0.695\pm 0.001$ (vs.~$0.815$ and $0.691$ for GR).
One-sided one-sample $t$-tests confirm that these improvements over GR are all statistically significant at $p<0.05$.

We further compare RQ-Reg with SOTA LTV prediction methods,
including Two-stage~\cite{drachen2018or}, MTL-MSE~\cite{ma2018entire}, ZILN~\cite{wang2019deep},
MDME~\cite{li2022billion}, MDAN~\cite{liu2024mdan}, and OptDist~\cite{weng2024optdist}.
As summarized in Table~\ref{tab:ltv}, RQ-Reg achieves competitive performance against all baselines on both datasets.
In particular, on the Gini~(+) and SRCC~(+) metrics,
which measure the ranking consistency between predicted values and ground-truth labels for positive samples,
RQ-Reg consistently outperforms every other SOTA method.
Specifically, compared with the second-best baseline, RQ-Reg yields relative improvements of $0.90\%$ and $7.24\%$ in Gini~(+) on Criteo-SSC and Kaggle respectively,
and relative improvements of $6.74\%$ and $0.22\%$ in SRCC~(+) on the two datasets respectively.
This indicates that RQ-Reg not only models the magnitude of LTV effectively,
but also preserves the relative ranking among positive samples more accurately.

\subsubsection{Ablation study.}
We conduct ablation experiments to study the impact of the key components in RQ-Reg.
We compare the full RQ-Reg with several variants in which some key component is modified,
using the watch-time prediction task for evaluation.

\textbf{Impact of training objectives.}
We evaluate two variants that ablate individual loss components:
\model{w/o $\mathcal{L}_\text{gen}$} removes the NLL loss, eliminating generative supervision over the predicted code sequence,
while \model{w/o $\mathcal{L}_\text{rnc}$} removes the RnC loss, the representation learning objective.
As shown in Table~\ref{tab:ab_main}, both variants exhibit consistent performance degradation.
The drop observed for \model{w/o $\mathcal{L}_\text{gen}$} confirms that sequential supervision over quantization codes
is crucial for capturing inter-step dependencies.
The degradation from removing $\mathcal{L}_\text{rnc}$ further demonstrates that learning high-quality latent representations for quantization codes
is essential for downstream prediction accuracy.

\textbf{Impact of regressors.}
Removing the regressor modules (\model{w/o $f_\text{reg}$ + $f_\text{err}$}) reduces the model to making predictions by taking the expectation over the conditional distribution,
\ie, $\hat{y}_l=\mathbb{E}_{p\left(q|\bs{x},q_{<l}\right)}[q]$.
The resulting performance drop in Table~\ref{tab:ab_main} suggests that direct regression aligns the model output with numerical targets more effectively
than the distribution-based expectation estimation.
Removing only the error regressor (\model{w/o $f_\text{err}$}) also leads to a consistent performance drop,
indicating that the error regressor is beneficial for refining the final predictions.

\textbf{Impact of scheduled sampling.}
We further investigate the effect of the scheduled sampling scheme.
Setting $p=1$ (\model{w/o SS~($p=1$)}) disables scheduled sampling by feeding the ground-truth code embedding $\bs{e}(q_l)$ into the LSTM at every step.
This causes a training-inference discrepancy and leads to a clear performance drop in Table~\ref{tab:ab_main}.
Conversely, setting $p=0$ feeds only the predicted code embedding $\hat{\bs{e}}_l$ into the LSTM throughout training,
which makes it harder for the model to learn reliable sequential dependencies.
Both extremes lead to degraded performance, highlighting the importance of gradually transitioning from teacher-forced embeddings to model-predicted ones
during training.

\begin{table}
  \caption{Comparison of different model backbones on watch-time prediction datasets}
  \begin{center}
  \begin{tabular}{lcccc}
  \toprule
  \multirow{2}{*}{Method} & \multicolumn{2}{c}{KuaiRec} & \multicolumn{2}{c}{CIKM16} \\
  \cmidrule{2-5}
   & MAE$^\downarrow$ & XAUC$^\uparrow$ & MAE$^\downarrow$ & XAUC$^\uparrow$ \\\midrule
  RNN & 3.210 & 0.613 & 0.815 & 0.694 \\
  LSTM & \textbf{3.189} & \textbf{0.616} & \textbf{0.811} & \textbf{0.695} \\
  Transformer & 3.204 & 0.614 & 0.829 & 0.694 \\
  \bottomrule
  \end{tabular}
  \end{center}
  \label{tab:ab_net}
\end{table}

\textbf{Choice of backbones.}
Table~\ref{tab:ab_net} compares the performance of different sequence modeling backbones:
vanilla RNN, LSTM, and Transformer~\cite{vaswani2017attention}.
Across both KuaiRec and CIKM16, the LSTM backbone consistently outperforms the vanilla RNN,
which we attribute to its gated memory mechanism that more effectively captures sequential dependencies.
Note that LSTM also outperforms the Transformer-based design in our experiments.
We believe that this is due to the limited scale of training data and the relatively small set of discretization codes in the watch-time prediction setting,
which may hinder the Transformer from fully exploiting its self-attention capacity.

\textbf{Comparison of quantization methods.}
We compare the RQ K-means approach against alternative quantization baselines, summarized in Table~\ref{tab:ab_code}.
Specifically, we include (1) \textbf{K-means}, which uses the same vocabulary size as RQ K-means,
and (2) \textbf{Dynamic quantile}, the heuristic discretization algorithm adopted by GR~\cite{ma2024generative}.
Experimental results show that RQ K-means outperforms both alternatives.
Moreover, since K-means quantization can be viewed as RQ K-means with $L=1$,
the performance gains observed as $L$ increases are well aligned with the coarse-to-fine nature of RQ-Reg,
in which additional levels progressively refine the predictions of earlier levels.
As reported in Table~\ref{tab:recon_err},
our RQ K-means achieves a substantially lower reconstruction error than the dynamic quantile method~\cite{ma2024generative} ($1.33\times 10^{-3}$ vs.~$3.28\times 10^{-3}$),
while using a shorter, fixed code length.
In addition, the inference complexity of our method scales linearly with the code length, in contrast to the quadratic scaling of GR~\cite{ma2024generative}.
RQ-Reg offers a clear computational advantage, achieving better reconstruction quality at a lower inference cost.

\begin{table}[!t]
  \caption{Comparison of different quantization methods on watch-time prediction datasets}
  \begin{center}
  \begin{tabular}{lcccc}
  \toprule
  \multirow{2}{*}{Method} & \multicolumn{2}{c}{KuaiRec} & \multicolumn{2}{c}{CIKM16} \\
  \cmidrule{2-5}
   & MAE$^\downarrow$ & XAUC$^\uparrow$ & MAE$^\downarrow$ & XAUC$^\uparrow$ \\\midrule
  K-means & 3.196 & 0.611 & 0.823 & 0.692 \\
  RQ K-means & \textbf{3.189} & \textbf{0.616} & \textbf{0.811} & \textbf{0.695} \\
  Dynamic quantile~\cite{ma2024generative} & 3.242 & 0.611 & 0.819 & 0.693 \\
  \bottomrule
  \end{tabular}
  \end{center}
  \label{tab:ab_code}
\end{table}

\begin{table}[!t]
  \caption{Reconstruction errors (in MAE) of different quantization methods on KuaiRec}
  \begin{center}
  \begin{tabular}{lcc}
  \toprule
  Method & Reconst.\ error & Code length\\
  \midrule
  K-means & $7.51\times 10^{-2}$ & 1 \\
  RQ K-means & $\mathbf{1.33\times 10^{-3}}$ & 3 \\
  Dynamic quantile~\cite{ma2024generative} & $3.28\times 10^{-3}$ & 3.21 (on avg.) \\
  \bottomrule
  \end{tabular}
  \end{center}
  \label{tab:recon_err}
\end{table}

\begin{figure}[!t]
  \centering
  \subfloat[Quantization error in MAE]{
    \includegraphics[width=3.8cm]{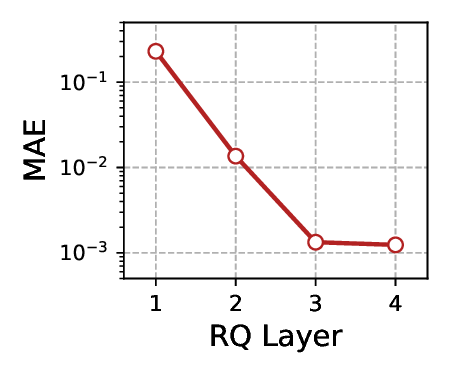}
    \label{subfig:layer_qe}
  }
  \subfloat[RQ-Reg prediction results]{
    \includegraphics[width=3.8cm]{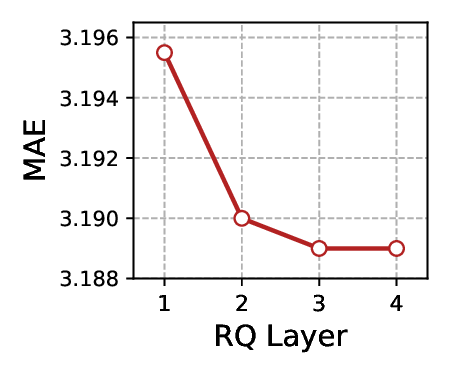}
    \label{subfig:layer_perf}
  }
  \caption{Results of RQ K-means quantization error (a) and RQ-Reg prediction (b) with different numbers of RQ levels $L$ on KuaiRec.}
  \label{fig:layer}
\end{figure}

\subsubsection{Effect of the number of RQ levels.}
We further investigate how the number of RQ levels $L$ affects both the quality of the learned coarse-to-fine code representation and the downstream prediction performance.
We vary $L$ from $1$ to $4$ and report (a) the final quantization error of the RQ K-means encoding,
and (b) the corresponding prediction MAE achieved by RQ-Reg on KuaiRec, as shown in Fig.~\ref{fig:layer}.

We observe in Fig.~\ref{subfig:layer_qe} that the reconstruction error decreases monotonically as $L$ grows,
confirming that RQ K-means effectively reduces the quantization error by adding levels and yields a progressively coarse-to-fine encoding of the target value.
Fig.~\ref{subfig:layer_perf} shows that the prediction MAE decreases as $L$ increases,
indicating that a more accurate quantization result directly translates into stronger predictive performance through the sequential refinement procedure of RQ-Reg.
Furthermore, we note that both curves begin to plateau with $L\ge 3$.
This suggests that the first few RQ levels contribute more critically to reducing quantization error and improving the prediction accuracy.

\subsubsection{Embedding visualization.}
\begin{figure}[!t]
  \centering
  \subfloat[RQ-Reg]{
    \includegraphics[width=3.9cm]{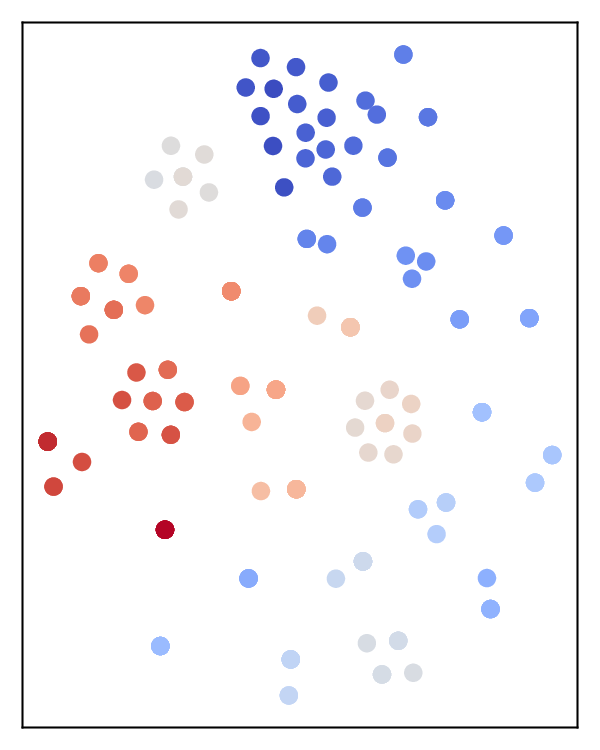}
    \label{subfig:emb_rnc}
  }
  \subfloat[w/o $\mathcal{L}_\text{rnc}$]{
    \includegraphics[width=3.9cm]{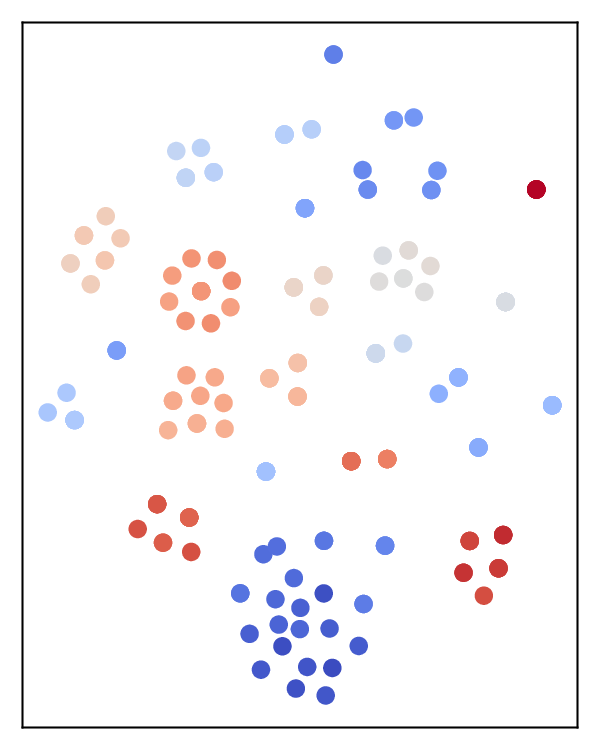}
    \label{subfig:emb_wo_rnc}
  }
  \caption{Learned representations of target values from different models on KuaiRec.
  Point color represents the magnitude of the corresponding target value, from low (cool) to high (warm).}
  \label{fig:emb}
\end{figure}

To better understand the impact of the continuous representation learning objective $\mathcal{L}_\text{rnc}$,
we investigate the learned embedding space from RQ-Reg model and its variant trained without $\mathcal{L}_\text{rnc}$ (denoted as \textbf{w/o $\mathcal{L}_\text{rnc}$}).
We first select a set of continuous target values and encode them into RQ code sequences using the RQ K-means method.
For a given target value $y$,
we map its RQ codes $(q_1,\cdots,q_L)$ to the embedding representations $\bs{e}=\left[\bs{e}(q_1);\cdots;\bs{e}(q_L)\right]$
using the embedding weights learned by RQ-Reg and \model{w/o $\mathcal{L}_\text{rnc}$} respectively.
We project these embeddings into two-dimensional space using t-SNE~\cite{maaten2008visualizing} and display the results.

Fig.~\ref{fig:emb} illustrates the t-SNE projections of target value embeddings on KuaiRec,
where each point corresponds to the representation of one target value, with its color denoting the target magnitude from low (cool) to high (warm).
For RQ-Reg (Fig.~\ref{subfig:emb_rnc}), labels with similar values exhibit close distances in the representation space,
indicating that the learned representation preserves the continuous rankings in the target space.
In contrast, the variant without $\mathcal{L}_\text{rnc}$ (Fig.~\ref{subfig:emb_wo_rnc}) produces scattered embeddings with noticeable outliers in the representation space.
These observations demonstrate that the continuous representation learning objective effectively enforces the geometrical consistency
between the representation and target spaces, which can further facilitate more accurate downstream predictions.

\subsection{Online Experiment}
\label{sub:online}
We deployed the proposed RQ-Reg in a large-scale online A/B test on Kuaishou,
which is a short-video recommendation platform serving over 400 million daily active users (DAU)
and delivering billions of video impressions per day.
The experiment targeted GMV prediction in the commercial advertising recommendation scenario.
In this advertising environment, the GMV distribution is highly clustered,
with sharp peaks at popular product price points.
This distributional pattern is primarily driven by the strong correlation between GMV and product pricing (see Fig.~\ref{fig:gmv}).

\subsubsection{Experiment setup.}
The advertising recommendation system at Kuaishou follows a two-stage framework.
In the first stage, a retrieval system generates a set of candidate items.
In the second stage, a ranking system selects the top items to display,
with the final ranking score jointly determined by two models:
one estimating purchase probability and the other forecasting potential GMV.
Our approach is integrated into the GMV prediction model of the second stage.
Specifically, the deployed RQ-Reg is configured with a $3\times 32$ RQ codebook and an LSTM backbone.
The production baseline for GMV prediction is based on WLR~\cite{covington2016deep},
which has been shown effective for continuous value prediction in large-scale recommendation,
\eg, watch-time prediction on YouTube~\cite{covington2016deep}.

We conducted a large-scale online A/B test, allocating 10\% of traffic to each of the two models (RQ-Reg and WLR),
which corresponds to approximately $40$ million unique users per model.
The experiment ran continuously for seven days.

\begin{table}
  \caption{Performance lift of RQ-Reg over the production baseline in the online short-video advertising scenario}
  \begin{center}
      \begin{tabular}{lcc}
          \toprule
          Setting & AUC$^\uparrow$ & ADVV$^\uparrow$ \\
          \midrule
          Overall & +0.12\% & +4.19\% \\
          Long-tail & +0.23\% & +4.76\% \\
         \bottomrule
      \end{tabular}
  \end{center}
  \label{tab:online}
\end{table}

\subsubsection{Results.}
We use two metrics to evaluate the online A/B test: AUC and advertiser value (ADVV).
AUC measures the ranking consistency between model scores and ground-truth purchasing behaviors over exposed impressions.
ADVV~\cite{chai2025longer} is a core business metric for advertising platforms that reflects the overall value generated for advertisers.
The online A/B test results are summarized in Table~\ref{tab:online}.
Overall, RQ-Reg achieves a notable improvement over the production baseline, with $+0.12\%$ AUC and $+4.19\%$ ADVV.
On impressions with long-tailed commodity prices, the gains become even more notable ($+0.23\%$ AUC and $+4.76\%$ ADVV),
demonstrating the effectiveness of RQ-Reg on highly complex value distributions.

\begin{figure}
  \centering
  \includegraphics[width=8.5cm]{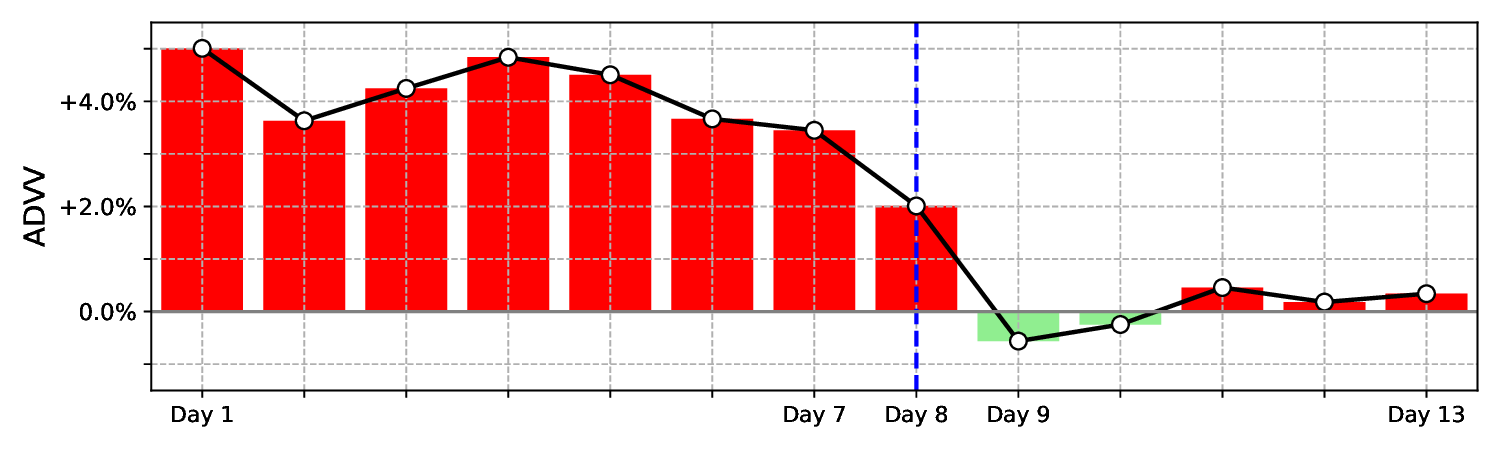}
  \caption{Daily ADVV gains during the online evaluation. The A/B test period spans days 1--7. On day 8 the experimental bucket was switched back to the production baseline,
  and the subsequent A/A test spans days 9--13.}
  \label{fig:online}
\end{figure}

Fig.~\ref{fig:online} reports the daily ADVV gains observed during the online evaluation.
During the A/B test period (days 1--7), the experimental bucket deployed RQ-Reg while the control bucket used the production baseline.
Throughout this period, RQ-Reg achieved consistent and statistically significant ADVV gains over the control.
On day 8, the serving model in the experimental bucket was switched to WLR, the production baseline.
The subsequent A/A test (days 9--13) showed a clear drop in ADVV compared to the A/B test period (days 1--7).
This sharp shift, observed immediately after the model swap, provides strong evidence of the effectiveness of RQ-Reg.

\section{Conclusion}
In this work, we propose RQ-Reg, a novel residual-quantization-based sequence learning framework for continuous value prediction.
By leveraging residual quantization (RQ), the method decomposes a continuous variable into a coarse-to-fine hierarchy of quantization codes,
enabling accurate recursive prediction while effectively mitigating quantization error.
We further introduce a representation learning objective that aligns the embedding space with the ranking of the target values,
allowing the model to learn continuous representations for quantization codes and improve prediction performance.

Extensive offline experiments and large-scale online evaluations
demonstrate that RQ-Reg performs favorably against state-of-the-art methods.
Notably, RQ-Reg exhibits strong capability in modeling complex value distributions and shows
remarkable generality and adaptability across diverse continuous prediction scenarios.
Future work includes investigating more effective quantization techniques for continuous value modeling in industrial settings,
and developing a deeper theoretical understanding of the proposed framework.

\appendix
\section{Additional Implementation Details}
\label{appen:detail}
We tune the regression loss weight $\alpha$ from $\{0.1,0.5,1.0, 2.0, 5.0, 10.0\}$,
and the RnC loss weight $\beta$ from $\{0.01,0.05,0.1, 0.2, 0.5, 1.0\}$,
and select the best, which is $\alpha=1.0$ and $\beta=0.1$, for the training objective function.
For the scheduled sampling scheme,
we search $k$ from $\{0.01,0.05, 0.1, 0.2, 0.5\}$
and $t_0$ from $\{0, 5, 10, 20, 30\}$,
and select $k=0.1$ and $t_0=10$ for all the datasets.
Due to the various target value ranges for different datasets,
we set $\delta=0.1$ in the Huber loss for Criteo-SSC, $\delta=10.0$ for CIKM16,
and adopt $\delta=2.0$ for the others.
We adopt the temperature $\tau=2.0$ and use negative $L_2$ norm
(\ie, $\text{sim}(\bs{u},\bs{v})=-\|\bs{u}-\bs{v}\|_2$) as the similarity measure for $\mathcal{L}_\text{rnc}$, following~\cite{zha2023rank}.

\section{Experiments on Hyperparameters}
We conduct experiments to investigate the sensitivity of RQ-Reg to its hyperparameters,
including the number of codes $K$ for each level,
weights $\alpha$, $\beta$ in the overall training objective,
and the hyperparameters $k$ and $t_0$ for the scheduled sampling scheme.

\begin{table}[h]
  \caption{Performance comparison with different settings of hyperparameter $K$}
  \begin{center}
  \begin{tabular}{lcccc}
  \toprule
  \multirow{2}{*}{Configuration} & \multicolumn{2}{c}{KuaiRec} & \multicolumn{2}{c}{CIKM16} \\
  \cmidrule{2-5}
    & MAE$^\downarrow$ & XAUC$^\uparrow$ & MAE$^\downarrow$ & XAUC$^\uparrow$ \\\midrule
  $K=16$ & 3.195 & 0.614 & 0.820 & 0.692 \\
  $K=32$ & 3.191 & 0.614 & 0.818 & 0.694 \\
  $K=48$ & 3.189 & 0.616 & 0.811 & 0.695 \\
  $K=64$ & 3.194 & 0.614 & - & - \\
  \bottomrule
  \end{tabular}
  \end{center}
  \label{tab:hp_codebook_size}
\end{table}

\begin{table}[h]
  \caption{Performance comparison with different settings of hyperparameters $\alpha$ and $\beta$}
  \begin{center}
  \begin{tabular}{lcccc}
  \toprule
  \multirow{2}{*}{Configuration} & \multicolumn{2}{c}{KuaiRec} & \multicolumn{2}{c}{CIKM16} \\
  \cmidrule{2-5}
    & MAE$^\downarrow$ & XAUC$^\uparrow$ & MAE$^\downarrow$ & XAUC$^\uparrow$ \\\midrule
    $\alpha = 0.1$ & 3.197 & 0.614 & 0.816 & 0.693 \\
    $\alpha = 0.5$ & 3.189 & 0.614 & 0.812 & 0.695 \\
    $\alpha = 1.0$ & 3.189 & 0.616 & 0.811 & 0.695 \\
    $\alpha = 2.0$ & 3.193 & 0.614 & 0.814 & 0.693 \\
    $\alpha = 5.0$ & 3.192 & 0.614 & 0.819 & 0.692 \\
    $\alpha = 10.0$& 3.200 & 0.613 & 0.820 & 0.691 \\
  \midrule
    $\beta = 0.01$& 3.189 & 0.615 & 0.815 & 0.694 \\
    $\beta = 0.05$& 3.190 & 0.615 & 0.814 & 0.694 \\
    $\beta = 0.1$ & 3.189 & 0.616 & 0.811 & 0.695 \\
    $\beta = 0.2$ & 3.193 & 0.615 & 0.812 & 0.694 \\
    $\beta = 0.5$ & 3.195 & 0.614 & 0.815 & 0.694 \\
    $\beta = 1.0$ & 3.195 & 0.614 & 0.818 & 0.693 \\
  \bottomrule
  \end{tabular}
  \end{center}
  \label{tab:hp_lambda}
\end{table}

\begin{table}[h]
    \caption{Performance comparison with different settings of hyperparameters $k$ and $t_0$}
    \begin{center}
    \begin{tabular}{lcccc}
    \toprule
    \multirow{2}{*}{Configuration} & \multicolumn{2}{c}{KuaiRec} & \multicolumn{2}{c}{CIKM16} \\
    \cmidrule{2-5}
      & MAE$^\downarrow$ & XAUC$^\uparrow$ & MAE$^\downarrow$ & XAUC$^\uparrow$ \\\midrule
      $k=0.01$ & 3.193 & 0.614 & 0.817 & 0.693 \\
      $k=0.05$ & 3.191 & 0.615 & 0.816 & 0.694 \\
      $k=0.1$ & 3.189 & 0.616 & 0.811 & 0.695 \\
      $k=0.2$ & 3.188 & 0.615 & 0.818 & 0.692 \\
      $k=0.5$ & 3.193 & 0.615 & 0.819 & 0.692 \\
    \midrule
      $t_0=0$ & 3.195 & 0.614 & 0.820 & 0.692 \\
      $t_0=5$ & 3.194 & 0.615 & 0.816 & 0.693 \\
      $t_0=10$ & 3.189 & 0.616 & 0.811 & 0.695 \\
      $t_0=20$ & 3.191 & 0.615 & 0.813 & 0.694 \\
      $t_0=30$ & 3.192 & 0.614 & 0.819 & 0.692 \\
    \bottomrule
    \end{tabular}
    \end{center}
    \label{tab:hp_sampling}
\end{table}

\textbf{Number of codes $K$.}
With the number of RQ levels fixed at $L=3$,
Table~\ref{tab:hp_codebook_size} reports the performance across different code numbers $K\in\{16,32,48,64\}$ for each level.
The results show that $K=48$ achieves the best performance on both datasets,
suggesting a favorable balance between classification accuracy and the reconstruction error introduced by discretization.
Note that RQ K-means with $K=64$ could not run successfully on the CIKM16 dataset, since the training data are insufficient to support such a large number of clusters.

\textbf{Loss function weights.}
The hyperparameters $\alpha$ and $\beta$ are the loss weights for regression and representation learning objectives.
Table~\ref{tab:hp_lambda} compares the performance with different settings of $\alpha$ and $\beta$.
We see that RQ-Reg is relatively insensitive to changes in $\alpha$ and $\beta$ on both datasets,
and the best performance is achieved when $\alpha=1.0$ and $\beta=0.1$.

\textbf{Scheduled sampling hyperparameters.}
We adopt the inverse sigmoid decay function defined in Eq.~(\ref{eq:inv_sigmoid}) to schedule the sampling rate,
where hyperparameters $k$ and $t_0$ control the shape of the decay curve.
The performances with different settings of $k$ and $t_0$ are shown in Table~\ref{tab:hp_sampling},
where $k=0.1$ and $t_0=10$ are selected as the default values.
When comparing the results in Table~\ref{tab:hp_sampling} and those without scheduled sampling, \ie, the variants \model{w/o SS~($p=0$)} and \model{w/o SS~($p=1$)} in Table~\ref{tab:ab_main},
we observe that the scheduled sampling scheme with various combinations of $k$ and $t_0$ can bring consistent performance gains.
This observation further illustrates the effectiveness of the scheduled sampling scheme.

\section*{GenAI Usage Disclosure}
During the preparation of this paper, the authors used Anthropic Claude Opus 4.7 as a writing assistant.
The tool was used solely to improve the writing quality, including grammar, wording, and clarity.
All scientific content, conclusions, code, experiments, and data analysis were produced and verified by the authors,
who take full responsibility for the entire content of the paper.

\bibliographystyle{ACM-Reference-Format}
\bibliography{reference}

\end{document}